\documentclass[twocolumn, twocolappendix]{aastex631}
\usepackage{enumitem}
\usepackage{xspace}
\usepackage{amsmath}
\usepackage{float}
\usepackage{booktabs}

\shorttitle{Baryons in the Darkest Sites of the Universe}
\shortauthors{Sharma et al.}

\begin{document}

\title{Baryons in the Darkest Sites of the Universe}

\correspondingauthor{Kritti Sharma}
\email{kritti@caltech.edu}

\author[0000-0002-4477-3625]{Kritti Sharma}
\affiliation{Cahill Center for Astronomy and Astrophysics, MC 249-17 California Institute of Technology, Pasadena CA 91125, USA.}

\author[0000-0002-7252-5485]{Vikram Ravi}
\affiliation{Cahill Center for Astronomy and Astrophysics, MC 249-17 California Institute of Technology, Pasadena CA 91125, USA.}
\affiliation{Owens Valley Radio Observatory, California Institute of Technology, Big Pine CA 93513, USA.}

\author[0000-0003-3312-909X]{Dhayaa Anbajagane}
\affiliation{Department of Astronomy and Astrophysics, University of Chicago, Chicago, IL 60637, USA}
\affiliation{Kavli Institute for Cosmological Physics, University of Chicago, Chicago, IL 60637, USA}

\author[0000-0002-1297-3673]{William R. Coulton}
\affiliation{Department of Physics, University of Oxford, Denys Wilkinson Building, Keble Road, Oxford OX1 3RH, United Kingdom}

\author[0000-0001-8356-2014]{Elisabeth Krause}
\affiliation{Department of Astronomy/Steward Observatory, University of Arizona, 933 North Cherry Avenue, Tucson, AZ 85721, USA}
\affiliation{Department of Physics, University of Arizona, 1118 E Fourth Street, Tucson, AZ 85721,
USA}

\author[0000-0001-5620-8554]{Nico Schuster}
\affiliation{Aix-Marseille Université, CNRS/IN2P3, CPPM, Marseille, France}

\author[0000-0002-6146-4437]{Alice Pisani}
\affiliation{Aix-Marseille Université, CNRS/IN2P3, CPPM, Marseille, France}
\affiliation{Department of Astrophysical Sciences, Princeton University, 4 Ivy Lane, Princeton, NJ 08544, USA.}

\author[0009-0008-5043-6220]{Samuel McCarty}
\affiliation{Center for Astrophysics, Harvard $\&$ Smithsonian, Cambridge, MA 02138-1516, USA.}

\author[0000-0002-7587-6352]{Liam Connor}
\affiliation{Center for Astrophysics, Harvard $\&$ Smithsonian, Cambridge, MA 02138-1516, USA.}

\author[0000-0003-4992-7854]{Simone Ferraro}
\affiliation{Lawrence Berkeley National Laboratory, 1 Cyclotron Road, Berkeley, CA 94720, USA}
\affiliation{Berkeley Center for Cosmological Physics, Department of Physics, University of California, Berkeley, CA 94720, USA}

\author[0000-0002-0876-2101]{Nico Hamaus}
\affiliation{Universitäts-Sternwarte München, Fakultät für Physik, LudwigMaximilians-Universität München, Scheinerstrasse 1, 81679
München, Germany}

\author[0000-0003-3714-2574]{Pranjal R. S.}
\affiliation{Department of Astronomy/Steward Observatory, University of Arizona, 933 North Cherry Avenue, Tucson, AZ 85721, USA}

\begin{abstract}

The pristine underdense patches of the Universe, cosmic voids, are powerful cosmological laboratories, uniquely sensitive to dark energy, modified gravity, and neutrino masses, yet their baryonic content remains uncharacterized. We present the first observational constraint on baryon underdensity in void interiors, exploiting the dispersion measures (DMs) of Fast Radio Bursts (FRBs) as tracers of the free electron column, independent of gas phase, temperature, and metallicity. By stacking 3,455 sightlines from CHIME/FRB with $\sim 15$~arcminute localizations on 1,288 SDSS BOSS voids over redshifts $0.2 < z < 0.7$, we measure a DM deficit toward void centers at $3.2\sigma$ significance, establishing that diffuse baryons inhabit the emptiest corners of the cosmic web at a suppressed level. The measured signal amplitude is consistent with an effective Universe model built directly from the observed galaxy underdensity in these voids, and a baryonic model calibrated to the FRB DM-redshift relation ($\alpha_v = 1.80 \pm 0.87$). A uniform-density void model yields an electron density contrast of $\delta_\mathrm{e,v} = -0.58 \pm 0.30$, implying a $\sim 60$\% underdensity of baryons in void interiors relative to the cosmic mean. Jointly interpreting our FRB measurement with existing stacks of the thermal Sunyaev-Zel'dovich effect on voids further constrains the mean void gas temperature to $T_\mathrm{e} \lesssim (1.1 \pm 0.7) \times 10^6$~K, pointing to a warm-hot diffuse phase, consistent with hydrodynamical simulation predictions. With forthcoming FRB (CHORD, DSA, SKA) and galaxy (DESI, LSST, Euclid, PFS-Subaru, SPHEREx, Roman) surveys, set to expand both samples by orders of magnitude, this approach opens a new window onto tomographic baryon mapping, with direct implications for feedback models governing gas expulsion into low-density environments, and for the use of cosmic voids to extract cosmological constraints.

\end{abstract}

\section{Introduction} \label{sec:introduction}

Cosmic voids~\citep{1982Natur.300..407Z, 1996Natur.380..603B, 1996ApJS..103....1B} -- the vast, underdense regions that fill the majority of the Universe's volume -- occupy a unique and increasingly prominent position in modern observational cosmology. Arising from the gravitational amplification of primordial density perturbations, voids form as matter drains away into surrounding filaments, walls, and clusters~\citep{1984MNRAS.206P...1I, 1993MNRAS.263..481V}, leaving behind expanses whose dynamics remain exquisitely close to the linear regime throughout cosmic history, retaining the memory of initial conditions~\citep{2009ApJ...701L..25S, 2019PhRvD..99l1304C, 2023NatCo..14.7523H}. This dynamical simplicity, which sets them fundamentally apart from the deeply non-linear environment of overdense cosmic structures~\citep{2014PhRvL.112y1302H, 2023JCAP...05..031S}, gives voids a suite of properties that make them extraordinarily powerful cosmological laboratories~\citep{2019BAAS...51c..40P, 2022LRR....25....6M, 2026A&ARv..34....1C}. The accelerated expansion shaping void sizes is governed by their dark-energy dominated interiors, rendering their size function uniquely sensitive to the dark-energy equation of state~\citep{2012MNRAS.426..440B, 2015PhRvD..92h3531P}. The low-density void environment also suppresses the screening mechanisms that hide modified-gravity signatures in denser regions, giving voids the capacity to probe deviations from General Relativity~\citep{2015PhR...568....1J, 2019LRR....22....1I}. Neutrinos, whose non-zero mass suppresses small-scale structure formation, have higher relative contribution to the mass budget in voids compared to other regions of the Universe, and the void sizes, spanning the free-streaming scales of neutrinos, directly constrain the sum of neutrino masses~\citep{2015JCAP...11..018M, 2019MNRAS.488.4413K, 2019JCAP...12..055S}.

The cosmological exploitation of voids has proceeded on multiple observational fronts~\citep{2022LRR....25....6M, 2026A&ARv..34....1C}. The void size function from SDSS BOSS has independently constrained the matter density, amplitude of density fluctuations, and the dark energy equation of state, with upper limits on the sum of neutrino masses~\citep{2023ApJ...953...46C, 2024ApJ...969...89T}. The void-galaxy correlations have been used to measure the growth rate of structure, with the derived posterior-distribution contours nearly orthogonal to those from overdensity-based probes, making void-overdensity combinations particularly powerful for breaking parameter degeneracies~\citep{2020JCAP...12..023H, 2020MNRAS.499.4140N, 2022MNRAS.513..186A, 2023MNRAS.523.6360W}. The weak gravitational lensing of background galaxies by voids has demonstrated sensitivity to modified gravity theories and the sum of neutrino masses~\citep{2014MNRAS.440.2922M, 2019MNRAS.490.3573F}. The average void shape serves as standard spheres for the Alcock-Paczy\'nski test~\citep{1995ApJ...452...25R, 2012ApJ...754..109L, 2012ApJ...761..187S, 2020JCAP...12..023H}. These measurements have been compounded by detections of cosmic microwave background (CMB) lensing~\citep{2017MNRAS.466.3364C} and the integrated Sachs-Wolfe (ISW) effect~\citep{2022MNRAS.515.4417K, 2025arXiv251016799D} from voids. With next-generation surveys, including the Dark Energy Spectroscopic Instrument~\citep[DESI; ][]{2024AJ....168...58D}, the Vera Rubin Observatory~\citep[LSST; ][]{2019ApJ...873..111I}, Euclid~\citep{2011arXiv1110.3193L}, the Prime Focus Spectrograph (PFS) survey on Subaru~\citep{2022arXiv220614908G}, SPHEREx~\citep{2020SPIE11443E..0IC}, and the Nancy Grace Roman Space Telescope~\citep{2019BAAS...51c.341D}, mapping millions of galaxies, voids are poised to move from being a complementary novelty to a central pillar of cosmological inference.

Despite the remarkable progress in exploiting the dark matter and galaxy content of voids for cosmology, comparatively little is known about their baryon content. Understanding baryon density in voids is important for several interconnected reasons. First, it directly impacts our ability to robustly identify $\lesssim 10$ Mpc-scale voids, as these smaller voids become progressively more polluted by galaxy formation physics with increasing strength of astrophysical feedback~\citep{2017MNRAS.470.4434P, 2024JCAP...08..065S}. Second, the relatively isolated environments of voids provide an ideal site for disentangling the role of mergers and large-scale environment in driving galaxy evolution~\citep{2019BAAS...51c..40P}. Third, the impact of the dynamical state of voids on the baryon cycle of resident halos is poorly understood: under hierarchical structure formation, halos in voids are expected to assemble more gradually, sustain efficient star formation, and host low-mass black holes which may thermally process their surrounding medium~\citep{2020MNRAS.493..899H, 2022MNRAS.511.2688R}. The impact of feedback on the lower-end of the void size function, as well as on void sub-structure, will become increasingly important as next-generation galaxy surveys achieve orders-of-magnitude higher tracer number densities.  

Observational access to the baryon content of voids has only recently begun to emerge. Measurements of the thermal Sunyaev-Zel'dovich (tSZ) effect, which probe the line-of-sight integrated electron pressure, have revealed a clear depletion of hot gas in these environments~\citep{2018PhRvD..97f3514A, 2024MNRAS.527.2663L}. Complementary insights from cosmological hydrodynamical simulations indicate that voids host a substantial baryon reservoir, where the exact amount and temperature of gas crucially depends on feedback strength~\citep{2016MNRAS.457.3024H, 2019MNRAS.486.3766M, 2024JCAP...08..065S}.

Fast radio bursts~\citep[FRBs; ][]{2022A&ARv..30....2P} have emerged as a transformative probe of the ionized baryon content of the Universe, uniquely suited to measure underdensities as well as overdensities along extragalactic sightlines. The dispersion measure (DM) of an FRB is a direct, unambiguous census of ionized baryons along the line of sight~\citep{2014ApJ...780L..33M}. The mean extragalactic DM, accumulated due to non-zero mean cosmic baryon density, increases monotonically with redshift, and is observationally well-established~\citep{2020Natur.581..391M}. FRB sightlines intersecting baryon overdensities in collapsed structures contribute excess DM~\citep{2025arXiv250608932W, 2025ApJ...993L..27H, 2025arXiv251102155T, 2026arXiv260121336S, 2026arXiv260422105S}, effectively up-scattering the extragalactic DM; the amplitude of this scatter reflects the degree of baryon clustering sculpted by the strength of feedback processes~\citep{2025ApJ...983...46M, 2025ApJ...989...81S, 2025arXiv250717742R, 2026arXiv260417162S}. Similarly, sightlines passing through baryon underdensities in cosmic voids produce a DM deficit relative to the cosmic mean, effectively down-scattering the extragalactic DM. The depth of this resulting ``DM cliff'' also directly traces the strength of feedback; stronger feedback evacuates more baryons from halos into voids, lowering the amplitude of underdensity relative cosmic mean, and hence, increasing the DM floor~\citep{2022MNRAS.509.4775J, 2024ApJ...965...57B}. 

FRBs have thus far been used extensively to study overdense structures -- filaments~\citep{2025NatAs...9.1226C}, galaxies~\citep{2025arXiv250608932W, 2025ApJ...993L..27H, 2026arXiv260121336S, 2026arXiv260422105S}, and clusters~\citep{2023ApJ...949L..26C, 2025arXiv251102155T, 2026arXiv260422105S} -- but their sensitivity to underdensities has not been exploited. Directly stacking FRBs on the positions of known cosmic voids isolates the void contribution to the total DM budget along each sightline and provides a direct measurement of the mean baryon underdensity inside voids, a quantity that has been theorized but never observationally accessed thus far. In this work, we take precisely this step. We cross-correlate the second catalog of the Canadian Hydrogen Intensity Mapping Experiment Fast Radio Burst (CHIME/FRB) sample of extragalactic DMs~\citep{2026ApJS..283...34C} with the SDSS BOSS cosmic void catalog~\citep{2017ApJ...835..161M} to directly measure the amplitude of baryon underdensity in voids imprinted on FRB DMs. This provides the first observational constraint on the mean baryon deficit inside voids. We additionally combine our measurement with tSZ-void stacks, which is sensitive to the product of electron density and temperature integrated along the sightline. The ratio of tSZ and DM signal therefore constrains the mean electron temperature inside voids. This combination offers a novel observational route to measure the density and thermal state of the diffuse gas inside cosmic voids, quantities intimately tied to the efficiency of feedback processes~\citep{2016MNRAS.457.3024H, 2017MNRAS.470.4434P, 2019MNRAS.486.3766M, 2022MNRAS.511.2688R}.

\begin{figure}[ht!]
\centering
\includegraphics[width=0.85\columnwidth]{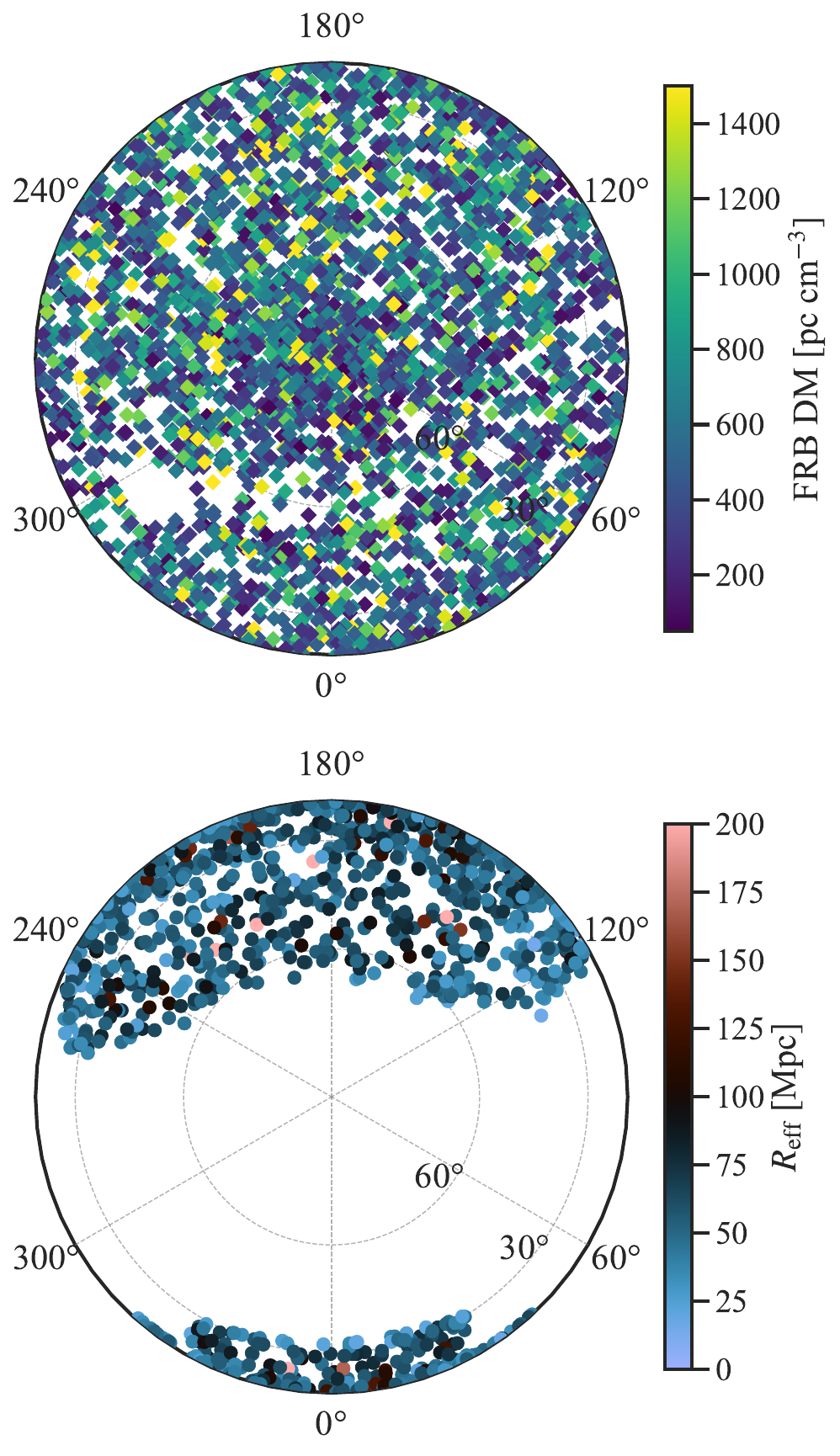}
\caption{Sky maps in an orthographic projection centered on the north celestial pole, showing the CHIME FRB sample~\citep{2026ApJS..283...34C}, colored by their extragalactic DMs in upper panel, alongside the SDSS void catalog~\citep{2017ApJ...835..161M}, colored by their effective void radius in lower panel. Dashed grid lines mark declinations of 30$^\circ$ and 60$^\circ$, and right ascension intervals of 60$^\circ$.}
\label{fig:maps}
\end{figure}

The remainder of this manuscript is organized as follows. In Section~\ref{sec:data}, we describe the FRB and void samples used in this analysis. Section~\ref{sec:modeling_baryon_underdensity} presents our theoretical framework for modeling the baryon underdensity in voids. We measure the DM-void cross-correlation in Section~\ref{sec:measured_anti_correlation}, conduct a suite of validation tests to assess the robustness of our measurement in Section~\ref{sec:robustness_tests}, and model it to quantify void baryon underdensity in Section~\ref{sec:baryon_underdensity}. We discuss the physical implications of our measurements and conclude in Section~\ref{sec:discussion_and_conclusion}. Throughout this work, we use the \citet{2020A&A...641A...6P} cosmology, which is consistent (within uncertainties) with the cosmology assumed by \citet{2017ApJ...835..161M} and \citet{2020JCAP...12..023H} in void finding procedures.

\section{Cosmic Voids and FRB Samples} \label{sec:data}

We use the void catalog of \citet{2017ApJ...835..161M} constructed by applying the ZOBOV void-finding algorithm~\citep{2008MNRAS.386.2101N} to the BOSS galaxy catalog from SDSS Data Release 12, spanning the redshift range $0.2 < z <0.7$ across $\sim$10,000~deg$^2$. ZOBOV performs a Voronoi tessellation of the galaxy distribution, using the resulting cell volumes as local density estimates, and groups adjacent cells into zones and voids via a watershed transform. Quality cuts are applied to only retain voids whose minimum Voronoi density falls below the mean sample density and whose statistical significance exceeds $2\sigma$, yielding 1,228 voids with effective radii, $R_\mathrm{eff} \sim 15-450~h^{-1}$Mpc. Void centers are defined as the inverse-density-weighted mean position of member galaxies (also known as the volume-weighted barycenter); this definition may not place the void center at the minimum density location. The effective radius $R_\mathrm{eff}$ is defined as the radius at which the enclosed mean density equals the mean cosmic density, such that voids are underdense for $R < R_\mathrm{eff}$. At $R \sim R_\mathrm{eff}$, a slight overdensity is expected due to the surrounding compensation wall of matter, before the profile saturates to the mean cosmic baryon density at larger radii.

For the FRB sample, we use DMs from the second CHIME/FRB catalog~\citep{2026ApJS..283...34C}. We retain events with well-measured sky positions, refined DM estimates from \textsc{fitburst} spectro-temporal modeling pipeline~\citep{2024ApJS..271...49F}, and restrict to a single burst per unique source. For cross-correlations, we use the NE2001-subtracted estimates of extragalactic DM~\citep{2002astro.ph..7156C, 2003astro.ph..1598C}. After these cuts, the sample comprises 3,455 sightlines, whose sky distribution is shown alongside the void catalog in Figure~\ref{fig:maps}. While our analysis does not directly use FRB redshift information, adopting the mean DM-redshift relation implies that selecting subsamples with extragalactic DM greater than 500, 750, and 1000~pc\,cm$^{-3}$, corresponds to median redshifts of $\sim$0.6, 0.8, and 1.0, respectively.

The localization uncertainty of this sample ranges from 0.2 to 32~arcminutes with a median of $\sim 15$~arcminutes. In Section~\ref{sec:robustness_tests}, we demonstrate that these large localization uncertainties are sufficient to probe underdensities within void interiors on scales of several tens of megaparsecs. FRBs are the best probe of baryons on these large scales, which are precisely where kinetic Sunyaev Zel'dovich (kSZ) effect suffers due to inevitable uncertainties from primary CMB anisotropies~\citep{2019PhRvD.100j3532M}. 

\section{Theoretical Model for Baryon Underdensity in Voids} \label{sec:modeling_baryon_underdensity}

By definition, cosmic voids occupy regimes where the foundational assumption of a halo model, namely, that all matter can be described as residing within virialized halos, breaks down. This motivates the use of alternative prescriptions, such as the effective universe method~\citep{2019PhRvD.100l3528J, 2020ApJ...889...89C}, which was originally developed in the context of void halo abundances to capture the dynamics of large-scale underdensities~\citep{2003MNRAS.344..715G, 2004ApJ...605....1G, 2015MNRAS.447.2683A, 2018PhRvD..97f3514A}. The key idea is that a spherically symmetric underdensity embedded in a homogeneous universe evolves, at any radius $r$ from its center, exactly like an independent Friedmann-Lema\^{i}tre-Robertson-Walker (FLRW) Universe with its own effective cosmological parameters, defined as follows~\citep{1947MNRAS.107..410B, 1993ppc..book.....P}.

Consider a population of voids characterized by a radial matter density contrast profile $\delta_\mathrm{m}(x=r/R_\mathrm{eff})$, where the matter underdensity is inferred directly from the galaxy underdensities of the parent sample as
\begin{equation}
    \delta_\mathrm{m}(x; z=0) = \dfrac{\delta_\mathrm{g}(x)}{b \, D(\langle z \rangle)/D(z=0)}.
    \label{eqn:matter_overdensity}
\end{equation}
Here, $D(z)$ is the linear growth factor~\citep{2025arXiv250907092S}, $b=2$ is the linear galaxy bias~\citep{2017MNRAS.469..787P} of this sample~\citep{2013MNRAS.432..743N}, and $\langle z \rangle$ is the mean redshift of the galaxy sample. Using Equation~\ref{eqn:matter_overdensity}, the mean enclosed density contrast $\Delta(x)$ is written as
\begin{equation}
    \Delta(x) = \dfrac{3}{x^3} \int\limits_0^x \mathrm{d}s \, s^2 \delta_\mathrm{m}(s).
    \label{eqn:Delta}
\end{equation}
The effective cosmological parameters are then directly defined using the mean enclosed density contrast as
\begin{equation}
    \Omega_\mathrm{m}(x) = \Omega_\mathrm{m} \dfrac{1+\Delta(x)}{\eta^2(x)}, \Omega_\Lambda (x) = \dfrac{\Omega_\Lambda}{\eta^2(x)}, H_0(x) = H_0 \eta(x),
    \label{eqn:effective_cosmology}
\end{equation}
with effective curvature $\Omega_\mathrm{k}(x) = 1 - \Omega_\mathrm{m}(x) - \Omega_\Lambda(x)$, where the local expansion rate factor $\eta(x)$ is determined by enforcing that the effective universe has the same age as the background cosmology. We solve for $\eta^2(x)$ numerically at each radial distance $x$ from the void center. Similarly, at each radius $x$, we also determine the effective redshift $z_\mathrm{eff}(x)$, defined as the redshift in the local effective cosmology that corresponds to the same cosmic time as the observed void redshift $z$ in the background cosmology. The amplitude of matter density fluctuations is also rescaled as
\begin{equation}
    \sigma_8(x) = \sigma_8 \dfrac{D (z_\mathrm{eff})}{D(z)}
    \label{eqn:effective_cosmology_s8}
\end{equation}
since the growth of density perturbations is slower inside the voids, as the cosmological constant dominates the local energy budget.

We compute the mean electron number density at radius $x$ by using the local effective cosmology $\{ \Omega_\mathrm{m}(x), \Omega_\Lambda(x), H_0(x), \sigma_8(x) \}$ as
\begin{equation}
    n_\mathrm{e}(x) = \int \mathrm{d}  M n(M|x) \int \mathrm{d}s \, 4 \pi s^2 n_\mathrm{e}(s, M),
    \label{eqn:void_electron_density_profile}
\end{equation}
where $n(M)$ is the halo mass function, and $n_\mathrm{e}(s, M)$ is the electron number density profile of a halo of mass $M$. This is computed using the \textsc{BCEmu} gas profile prescription, as implemented in \textsc{BaryonForge}\footnote{\url{https://github.com/DhayaaAnbajagane/BaryonForge}}~\citep{2024OJAp....7E.108A}, with feedback parameters fixed to median of posteriors inferred from DM-redshift relation in \citet{2026arXiv260417162S}. Since $\Omega_\mathrm{m}(x)$ and $\sigma_8(x)$ are reduced inside the void, the halo mass function is shifted towards lower mass halos, thus suppressing the contribution of massive halos to the total electron budget compared to an average density region. The resulting electron density profile monotonically increases outward and asymptotes to the cosmic mean, $n_\mathrm{e}(x \rightarrow \infty) = \bar{n}_\mathrm{e}$, reflecting the void's underdensity relative to the cosmic mean. The three-dimensional electron underdensity profile is then defined as
\begin{equation}
    \Delta n_\mathrm{e}(x) = n_\mathrm{e}(x) - n_\mathrm{e}(x \rightarrow \infty).
    \label{eqn:void_electron_density}
\end{equation}
The underdensity profile is projected along the line of sight using an Abel integral~\citep{Abel} to obtain the predicted DM underdensity at projected impact parameter $b$ (normalized by the void effective size $R_\mathrm{eff}$, so $b\leq1$),
\begin{equation}
    \Delta \mathrm{DM}(b) = \langle R_\mathrm{eff} \rangle \int\limits_b^{1} \dfrac{\Delta n_\mathrm{e}(x) \, 2x \, \mathrm{d}x}{ \sqrt{x^2 - b^2}},
    \label{eqn:void_DM_profile_effective}
\end{equation}
where we assume self-similarity of void profiles and approximate the projection using the mean effective void radius $\langle R_\mathrm{eff} \rangle$.

In addition to this effective universe approach, we also consider a simpler uniform-density void model, which assumes a constant fractional electron density contrast $\delta_\mathrm{e,v}$ throughout the void interior, so that
\begin{equation}
    n_\mathrm{e}(r) = \bar{n}_\mathrm{e}(z) (1+\delta_\mathrm{e,v}), \, r \leq R_\mathrm{eff}.
    \label{eqn:delta_v}
\end{equation}
The projected DM profile for this model is then
\begin{equation}
    \Delta \mathrm{DM}(b) = 2\bar{n}_\mathrm{e}(z) \delta_\mathrm{e,v} \sqrt{R_\mathrm{eff}^2(1-b^2)},
    \label{eqn:void_DM_profile_uniform}
\end{equation}
where the amplitude is directly proportional to $\delta_\mathrm{e,v}$. While less physically motivated than the effective universe approach, this model provides a direct, model-independent handle on the mean fractional baryon underdensity in void interiors, and its single free parameter $\delta_\mathrm{e,v}$ can be straightforwardly compared with constraints from other probes, such as tSZ effect.

\section{Measurement of Anti-Correlation} \label{sec:measured_anti_correlation}

We measure the physical-space correlation function $\hat{\xi}^{v\mathcal{D}}(R)$ between the void catalog and the FRB DM field using \textsc{treecorr} package~\citep{2015ascl.soft08007J}, following the methodology described in \citet[][Equation~34]{2026arXiv260422105S}. Voids act as transmissive dispersers weighted by their effective radii, and the estimator quantifies the mean DM excess or deficit around void centers relative to random positions at projected physical separation $R$. The correlation is measured in five logarithmically spaced bins spanning $10 \leq R \leq 100$~Mpc. We construct a joint data vector by concatenating measurements across three nested FRB subsets, selected by applying lower DM threshold cuts of \{500, 750, 1000\}~pc\,cm$^{-3}$, which progressively select more distant and cosmologically informative sightlines. Since these subsets overlap by construction, the correlations between measurements at different DM thresholds are fully captured in the joint jackknife covariance matrix, estimated over 40 spatial patches with the \citet{2007A&A...464..399H} correction applied to de-bias the inverse covariance matrix. When conducting parameter inference, we additionally inflate our inferred parameter variance by the \citet{2013PhRvD..88f3537D} correction factor. The combined signal-to-noise is computed from the joint data vector following \citet{2022PhRvD.105j3537S}. We refer the reader to \citet{2026arXiv260422105S} for a detailed description of the covariance estimation and signal-to-noise methodology. For completeness, we also performed the measurement using three non-overlapping differential bins, finding that the SNR ($2.5\sigma$) is dominated by FRBs with extragalactic DM $\geq 1000$~pc\,cm$^{-3}$.

\begin{figure}
\centering
\includegraphics[width=\columnwidth]{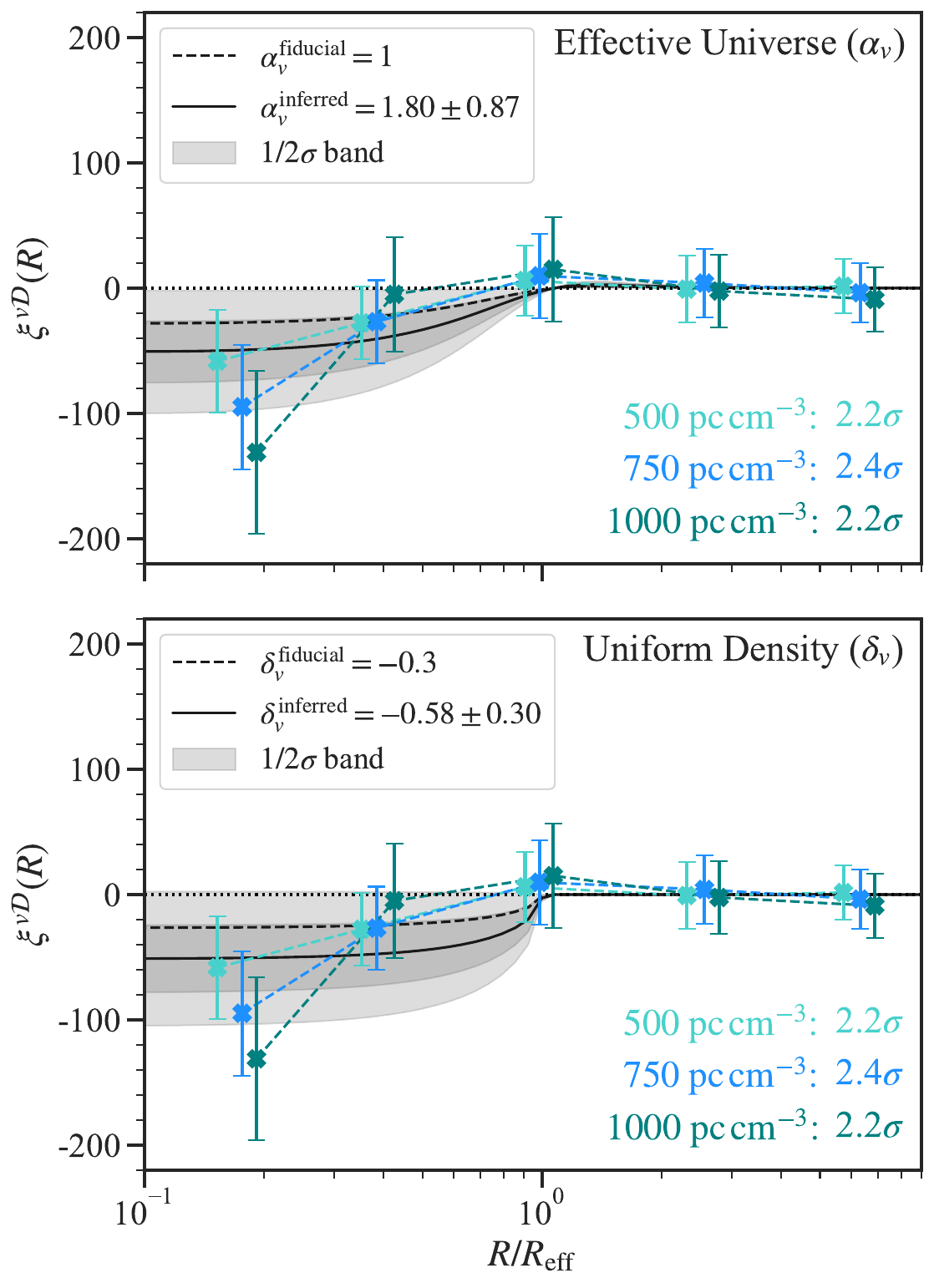}
\caption{The measured anti-correlation between FRB DMs and void positions at $3.2\sigma$ statistical significance. The colors correspond to three DM thresholds used to construct the nested FRB subsets for cross-correlations, defined as DM $\geq$ DM$_\mathrm{cut}$, where DM$_\mathrm{cut} \in $ \{500, 750, 1000\}~pc\,cm$^{-3}$. These subsets are sensitive to broadly different redshift ranges. We model the observed signal with an effective universe approach (upper panel), and a uniform density approach (lower panel). The amplitude of the measured signal is consistent with theoretical prediction from effective universe model ($\alpha_v = 1.28 \pm 0.57$), and reflects a $\sim 60$\% underdensity of baryons in void interiors compared to the surrounding medium ($\delta_v = -0.51 \pm 0.27$).}
\label{fig:voids_measurement}
\end{figure}

\begin{figure}
\centering
\includegraphics[width=\columnwidth]{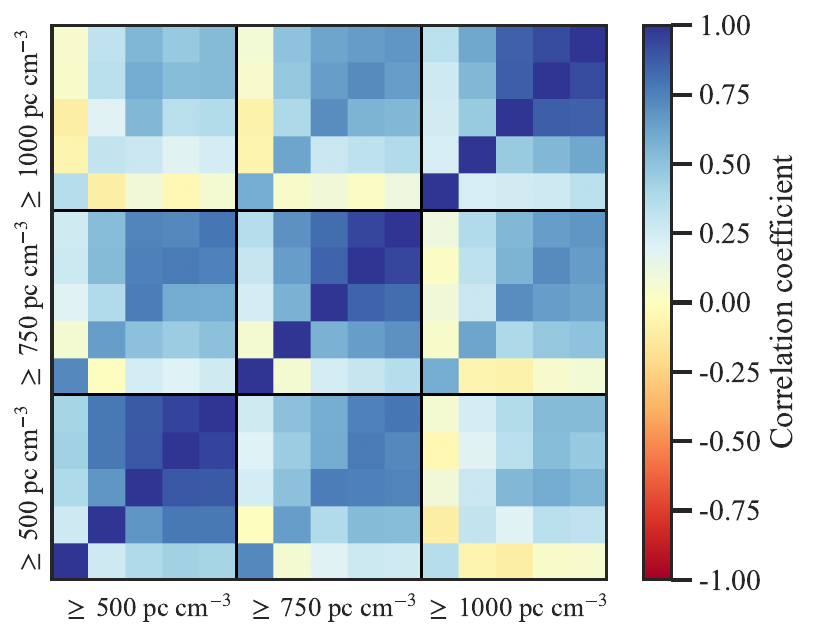}
\caption{Correlation matrix from the jackknife covariance between angular bins, evaluated jointly for the three DM threshold samples. Each entry represents the Pearson correlation coefficient $C_{ij}/\sqrt{C_{ii}C_{jj}}$. The block structure highlights correlations between separation bins at fixed DM threshold, while the off-diagonal blocks encode correlations between different DM cuts. These inter-sample correlations arise naturally because samples with higher DM thresholds are nested within those defined by lower thresholds.}
\label{fig:voids_correlation_matrix}
\end{figure}

\begin{figure*}[ht!]
\centering
\includegraphics[width=\textwidth]{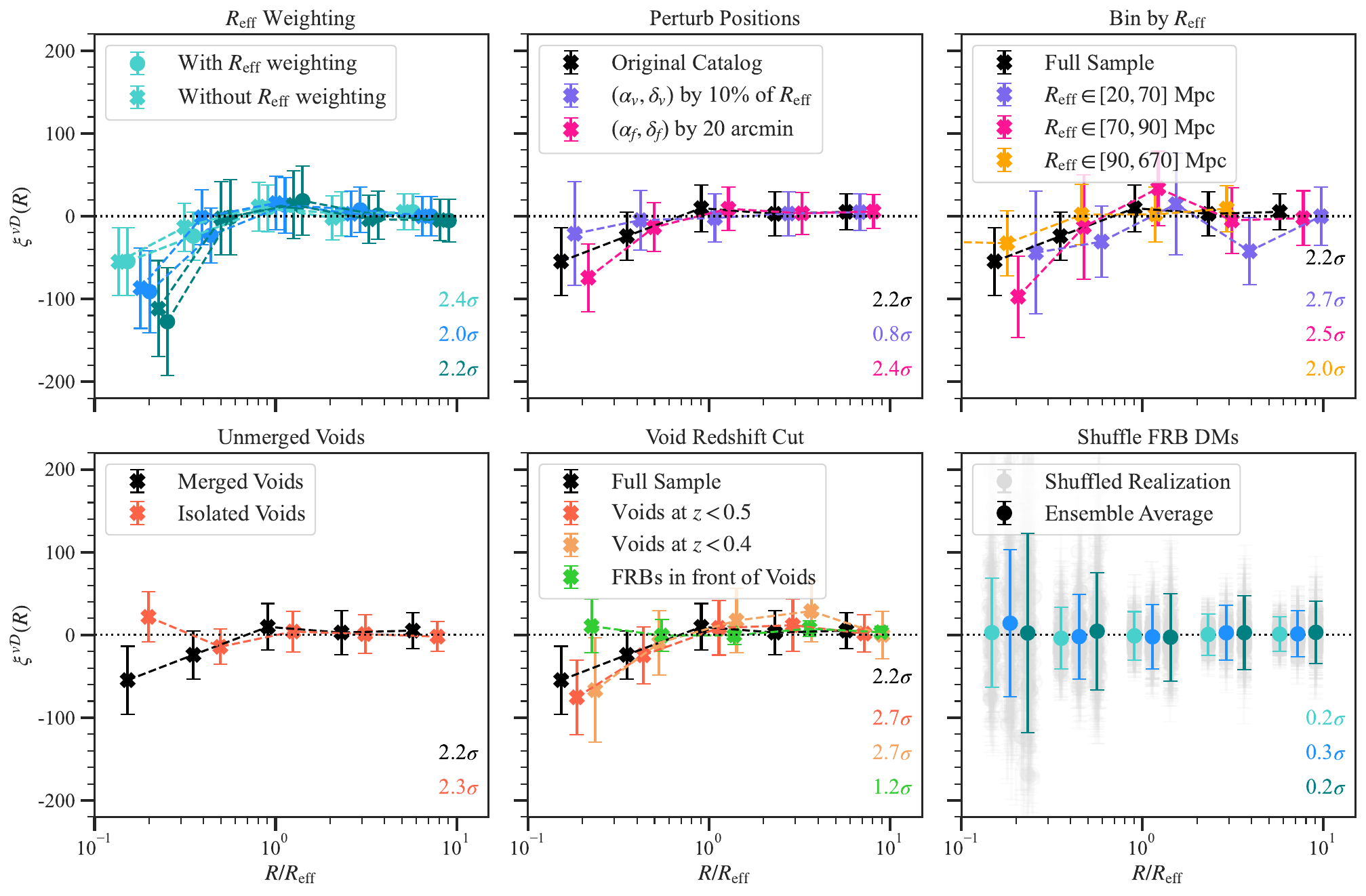}
\caption{Validation tests to assess the robustness of our measurement. All, except the first and last few tests, are conducted using the FRB sample thesholded at 500~pc\,cm$^{-3}$ for visual clarity. The upper-left panel shows the sensitivity of measurement to void size weighting used in our cross-correlation estimator; the $R_\mathrm{eff}$ weighting impacts the measurement at $0.2-0.4\sigma$ level. The upper-middle panel verifies the sensitivity of measured signal to positional uncertainty of FRBs and voids, where the uncertainty from void center measurement may be a dominant systematic. The upper-right panel evaluates the contribution to observed signal from small and large voids; all void sizes are contributing to the signal, including the ones with sizes well beyond the mean tracer (BOSS galaxies) separation. The bottom-left panel compares the measured DM deficit to that obtained using a larger void catalog constructed from BOSS galaxies without merging smaller voids, in which case the underdensity vanishes, indicating the importance of merging smaller associated Voronoi cells in watershed algorithms. The bottom-middle panel compares the signal strength when redshift cuts are applied to ensure voids lie in front of the FRB sources; the signal remains persistent and consistent within the error bars, and vanishes when FRBs are placed in front of the voids. The bottom-right panel tests the null hypothesis, where randomization of FRB DMs while fixing the void and FRB positions de-correlates the signal.}
\label{fig:voids_validation_tests}
\end{figure*}

We present the cross-correlation measurements between the FRB DM field and the SDSS void catalog in Figure~\ref{fig:voids_measurement}, and the corresponding correlation matrix in Figure~\ref{fig:voids_correlation_matrix}. Cosmic voids, as the most under-dense regions of the large-scale structure, represent a physically distinct regime from the overdense environments probed by other tracers presented in \citet{2025arXiv250608932W} and \citet{2026arXiv260422105S}. Rather than a positive DM excess, we expect a negative cross-correlation signal (a DM deficit around void centers) since FRB sightlines passing through under-dense void interiors accumulate less free electron column than the cosmic mean. Consistent with this physical picture, the void catalog yields an anti-correlation across all DM thresholds, with the signal recovering toward the cosmic mean near $\sim R_\mathrm{eff}$ as the sightlines exit void interiors. The per-DM-cut measurements are $2.2\sigma,~2.4\sigma$, and $2.2\sigma$, combining to a joint SNR of $3.2\sigma$, providing direct observational evidence that the diffuse baryon distribution faithfully traces the under-dense skeleton of the cosmic web. The void measurements are thus a natural complement to the overdensity measurements of \citet{2025arXiv250608932W} and \citet{2026arXiv260422105S}.

\section{Robustness Tests}\label{sec:robustness_tests}

We assess the robustness of our measurements through a series of validation tests designed to understand the origin of the signal and to quantify the systematic effects that could otherwise generate a spurious signal. 

We begin by investigating the impact of including $R_\mathrm{eff}$ weighting in our cross-correlation estimator. In the top-left panel of Figure~\ref{fig:voids_validation_tests}, we show the measurements with and without this weighting. We find that incorporating $R_\mathrm{eff}$ weighting increases the SNR at $0.2-0.4\sigma$ level.

Next, we verify whether the measured signal is an artifact of positional uncertainties in either the void center locations or the FRB localizations. We perform these tests using the FRB sample defined with a lower DM threshold of 500~pc\,cm$^{-3}$ and show the measurements in the upper-middle panel of Figure~\ref{fig:voids_validation_tests}. The void center in \citet{2017ApJ...835..161M} is defined as the volume-weighted barycenter of all the Voronoi cells constituting the void. Perturbing each void center by a random displacement of 10\% of its $R_\mathrm{eff}$ (corresponding to angular displacement in the range of $\sim 5$~arcminute to $\sim 3$~deg) and recomputing the cross-correlation results in a reduced measurement significance of $0.8\sigma$ (relative to original measurement SNR of $2.2\sigma$), indicating that the signal is sensitive to errors in void center determination (equivalent to halo mis-centering uncertainties in cluster science~\citep{2012ApJ...757....2G}). While a volume-weighted center definition is generically robust, it is still possible that the center does not fall in the exact most underdense part of the void. In contrast, perturbing each FRB position by a random displacement of $20$~arcmin, comparable to the CHIME/FRB localization uncertainty, results in only a modest change in measurement significance from $2.2\sigma$ to $2.4\sigma$, confirming that the CHIME/FRB localization precision is sufficient for conducting this measurement. This reflects the fact that the projected area associated with CHIME localization uncertainties at void redshifts remains much smaller than the characteristic size of the voids themselves.

We further evaluate the susceptibility of the observed signal to spurious voids at the small and large-ends of the void size function by splitting the observed voids into three bins of equal sizes with $R_\mathrm{eff} \in [20, 70]$~Mpc, $R_\mathrm{eff} \in [70, 90]$~Mpc, and $R_\mathrm{eff} \in [90, 670]$~Mpc, and cross-correlate each subset with the FRB sample constructed at a DM threshold of $500$~pc\,cm$^{-3}$. While the measurement itself is performed in fixed physical distance bins, as described in Section~\ref{sec:measured_anti_correlation}, for visualization in the figures, we express the radial distance in relative units, where we rescale the x-axis by dividing by the median $R_{\mathrm{eff}}$ of the corresponding void subset (after cuts) used in each case. This allows us to present the results in units of $R/R_{\mathrm{eff}}$ while keeping the underlying binning in physical distances. As illustrated in upper-right panel of Figure~\ref{fig:voids_validation_tests}, all void sizes are contributing to the signal, with contributions consistent within $1\sigma$ uncertainties. We have also tested the robustness of the measured signal to the exclusion of voids with radii below twice and thrice the mean tracer separation ($\approx 18$~Mpc\,$h^{-1}$). We find that the joint SNR decreases only modestly from $3.2\sigma$ to $3.0\sigma$, with no visually significant change in the measured correlation function, indicating that our results are not driven by small, potentially less reliable voids. A better evaluation of the behavior of voids with sizes smaller than the mean tracer separation could be performed in the future with larger samples. In the bottom-left panel of Figure~\ref{fig:voids_validation_tests}, we present the importance of merging smaller isolated voids~\citep{2020JCAP...12..023H} to their parent voids~\citep{2017ApJ...835..161M} before conducting such stacking analyses.

Another useful sanity check is to place FRBs in front of voids and examine how the signal behaves. To this end, we impose a DM cut of $\leq 400$~pc\,cm$^{-3}$ and select voids at redshifts $z \geq 0.5$. While this is not a perfect test, since the scatter in the DM-redshift relation still allows some FRBs to lie beyond $z = 0.5$, we nevertheless expect the correlation to be significantly suppressed. As shown in the bottom-middle panel of Figure~\ref{fig:voids_validation_tests}, the measured signal is consistent with the null expectation. We further test the robustness of this result by imposing cuts on void redshifts, requiring $z < 0.4$ and $z < 0.5$, while imposing a DM cut of $\geq 500$~pc\,cm$^{-3}$, to ensure that sightlines which do not intersect higher-redshift voids under a DM threshold of 500~pc\,cm$^{-3}$ do not contribute significant additional noise to the observed signal. Under these cuts, we find that the SNR increases modestly from $2.2\sigma$ to $2.7\sigma$, reinforcing the robustness of our measurement.

Finally, we randomly shuffle the FRB DMs while keeping their on-sky positions fixed, de-correlating the baryon distribution from the large-scale structure to compute the null expectation. The results from 100 such shuffles, presented in the bottom-right panel of Figure~\ref{fig:voids_validation_tests}, are consistent with zero signal across all radial bins and DM thresholds, confirming that the measured anti-correlation is not a statistical artifact of the small sample size. Taken together, these validation tests establish the robustness of the $3.2\sigma$ measurement and support its physical interpretation as a genuine baryon underdensity in the interior of cosmic voids.

\section{Baryon Underdensity in Voids} \label{sec:baryon_underdensity}

The $3.2\sigma$ anti-correlation measurement in Section~\ref{sec:measured_anti_correlation} constitutes the first direct observational evidence for a baryon deficit in the interiors of cosmic voids. We now quantify this deficit by fitting the measured cross-correlation signal against the two complementary theoretical models described in Section~\ref{sec:modeling_baryon_underdensity}, assuming a Gaussian likelihood (see Equations 41 and 42 of \citet{2026arXiv260422105S}). The results from both of these models are displayed in Figure~\ref{fig:voids_measurement}.

In the effective universe framework (Equation~\ref{eqn:void_DM_profile_effective}), the baryon underdensity inside a void is not a free parameter, but instead follows deterministically from the observed galaxy underdensity in the void catalog, modulated by a single amplitude scaling parameter $\alpha_v$ that encodes the fidelity of our baryonic prescription and the calibration of the DM-redshift relation. We have verified that our calculation reproduces the galaxy underdensity profile of voids from \citet{2017ApJ...835..161M}, which we use for the effective universe calculations. Fitting the cross-correlation signal jointly over radial bins $R \leq R_\mathrm{eff}$, we recover $\alpha_v = 1.80 \pm 0.87$, consistent with the fiducial prediction $\alpha_v=1$ at $\lesssim 1\sigma$ level. This agreement indicates that the observed DM deficit is quantitatively consistent with a universe in which the baryon distribution is described by the best-fit gas profile model of \citet{2026arXiv260417162S}, embedded in an effective local cosmology -- one where the reduced matter density and suppressed growth rate inside the void shift the halo mass function toward lower masses, depleting the electron budget relative to the cosmic mean.

To extract a more direct, model-independent characterization of the baryon content, we additionally fit the simpler uniform-density void model (Equation~\ref{eqn:void_DM_profile_uniform}), which parametrizes the mean fractional electron density contrast inside voids as a single constant $\delta_\mathrm{e,v}$. This model makes no assumption about the radial profile or physical origin of the underdensity. Fitting over the same radial range, we find $\delta_\mathrm{e,v} = -0.58 \pm 0.30$, a $1.4\sigma$ preference for an electron underdensity. The implied ratio of the mean electron number density inside these voids to the cosmic mean is $n_\mathrm{e,v}/\bar{n}_\mathrm{e} = 1 + \delta_\mathrm{e,v} = 0.42 \pm 0.30$, indicating that the free electron column through void interiors is approximately half the cosmic mean. This constitutes the first direct observational constraint on the mean baryon underdensity in voids.

Our measurement is in good agreement with upper limits derived from the tSZ effect by \citet{2024MNRAS.527.2663L}, who found $n_\mathrm{e,v}/\bar{n}_\mathrm{e} \lesssim 0.73$ (Atacama Cosmology Telescope, ACT) and $n_\mathrm{e,v}/\bar{n}_\mathrm{e} \lesssim 0.49$ (Planck) at 95\% confidence from constraints on $\delta_\mathrm{e,v} T_\mathrm{e}$ combined with hydrodynamical simulation-motivated priors on the void gas temperature. Stacking the galaxies tracing the voids of the \citet{2017ApJ...835..161M} sample at void positions, we measure a central galaxy density which is $\sim 20$\% of the mean sample density, corresponding to a galaxy underdensity of $\sim 80$\%. Under the uniform density model, this translates to $\delta_\mathrm{g,v} \sim -0.39$ and $\delta_\mathrm{m,v} \sim -0.26$, consistent with our measured $\delta_\mathrm{e,v}$ within $\sim 1\sigma$ measurement uncertainty. This suggests that baryons approximately trace the sparse halo population inside voids, though the uncertainties are too large to distinguish between scenarios where baryons trace dark matter vs mild baryon expulsion. Tighter constraints from forthcoming surveys will be needed to adjudicate between these possibilities.

Since DM and tSZ Compton-$y$ probe different moments of the electron density ($n_\mathrm{e}$): the DM traces $\int n_\mathrm{e} \, \mathrm{d}l$, while the $y$-parameter traces $\int n_\mathrm{e} T_\mathrm{e} \, \mathrm{d}l$, their ratio directly constrains the mean electron temperature ($T_\mathrm{e}$) inside voids, independent of assumptions about metallicity. Combining our measurement of $\delta_\mathrm{e,v} = -0.58 \pm 0.30$ with the tSZ void stacks of \citet{2018PhRvD..97f3514A} and \citet{2024MNRAS.527.2663L}, which yield $\delta_\mathrm{e,v} T_\mathrm{e} = -(6.5 \pm 2.3) \times 10^5$~K (ACT) and $\delta_\mathrm{e,v} T_\mathrm{e} = -(8.6 \pm 2.1) \times 10^5$~K (Planck), we derive $T_\mathrm{e} = (1.1 \pm 0.7) \times 10^6$~K and $T_\mathrm{e} = (1.5 \pm 0.9) \times 10^6$~K, respectively. Comparing with gas temperature measurements of BOSS galaxies inferred by combining kSZ and tSZ effect data~\citep{2021PhRvD.103f3513S}, unsurprisingly, we find our measurement in voids to be an order of smaller. These temperatures point to a warm-hot diffuse gas phase in void interiors, consistent with predictions from hydrodynamical simulations~\citep{2016MNRAS.457.3024H, 2019MNRAS.486.3766M}. We note that the true volume-averaged gas temperature is likely lower, since the dependence of the tSZ signal on temperature renders it insensitive to the cooler phases that may dominate the baryon mass budget in voids.

\section{Discussion and Conclusion} \label{sec:discussion_and_conclusion}

Cosmic voids occupy a large fraction of the Universe's volume and represent its most extreme underdensities. However, the baryon content of voids is the least observationally constrained of all regimes of large-scale structure. The first detections of a tSZ decrement toward cosmic voids~\citep{2018PhRvD..97f3514A, 2024MNRAS.527.2663L} established that voids are under-pressured relative to the cosmic mean, in qualitative agreement with expectations from the environmental modulation of halo abundances. The present work extends this picture to the free electron column directly. By cross-correlating 3,455 CHIME/FRB sightlines with 1,288 SDSS BOSS voids over $0.2 < z < 0.7$, we measure a $3.2\sigma$ DM deficit toward void centers that is negative at small projected separations and recovers to zero at $\sim R_\mathrm{eff}$ -- the characteristic signature of a baryon underdensity confined to void interiors. The constraints derived in Section~\ref{sec:baryon_underdensity} from two complementary models -- $\alpha_v = 1.80 \pm 0.87$ from the effective universe framework, and $\delta_\mathrm{e,v} = -0.58 \pm 0.30$ from the uniform-density parameterization -- place the first direct observational bounds on the mean baryon density in void interiors. Our void catalog subtends approximately 27\% of the sky within the survey footprints considered, so the measured anti-correlation is not driven by a small, localized region, but instead represents a statistical property of the underdense skeleton of cosmic large-scale structure.

The combination of our FRB DM measurement with existing tSZ void stacks, as described in Section~\ref{sec:baryon_underdensity}, is particularly informative for testing models of non-local heating in the intergalactic medium. Some models predict that blazar heating or AGN feedback could generate an inverted temperature-density relation in voids~\citep{2008MNRAS.386.1131B, 2012ApJ...752...23C}. The electron temperatures we derive, $T_\mathrm{e} \lesssim (1.1 \pm 0.7) \times 10^6$~K, are consistent with the baryon make up in voids comprising a warm-hot phase that contributes to the tSZ signal, as predicted by hydrodynamical simulations~\citep{2016MNRAS.457.3024H, 2019MNRAS.486.3766M}. These temperature estimates, however, carry significant uncertainty at present. Future FRB datasets from the Canadian Hydrogen Observatory and Radio-transient Detector~\citep[CHORD;][]{2019clrp.2020...28V}, the Deep Synoptic Array~\citep[DSA;][]{2019BAAS...51g.255H}, and the Square Kilometer Array~\citep[SKA;][]{2004NewAR..48..979C}, with larger sky coverage, higher on-sky number densities, and precise redshift information to separate out the DM deficit/excess from the mean DM contribution, will be needed to determine whether void heating by non-local feedback is observationally distinguishable from the more standard warm-hot intergalactic medium scenario.

The present measurement also has implications for the use of voids as cosmological probes. The $\sim 60$\% baryon underdensity we measure is large enough to be potentially relevant for small-scale void finding: astrophysical feedback processes that expel gas into void interiors could bias the identification of voids from galaxy catalogs, particularly for the smallest voids, whose size function is sensitive to the tracer density and its coupling to baryonic physics~\citep{2017MNRAS.470.4434P, 2024JCAP...08..065S}. As next-generation galaxy surveys, such as DESI, LSST, Euclid, PFS-Subaru, SPHEREx, and Roman achieve higher tracer densities, the interplay between baryonic feedback and void statistics will become an increasingly important systematic to characterize and control. Furthermore, the combination of FRB DMs with matter tracers may allow tests of feedback models via comparisons of baryon and dark matter densities in the same voids.

Looking ahead, the methodology introduced here -- stacking FRB DMs on void positions -- naturally scales with the rapid growth of both FRB and galaxy survey datasets. Forthcoming FRB experiments, including CHORD, DSA, and SKA, are expected to deliver catalogs of order $10^4 - 10^6$ well-localized bursts, expanding the FRB sample by orders of magnitude beyond the current CHIME catalog. Combined with increasingly precise void catalogs from DESI, LSST, Euclid, PFS-Subaru, SPHEREx, and Roman, and higher-resolution tSZ maps from the Simons Observatory~\citep{2024ApJS..274...33G}, this approach will enable tomographic mapping of baryon underdensities and gas temperature. These datasets will tighten constraints on the warm-hot intergalactic medium and on feedback models that regulate gas expulsion into low density environments, establishing void interiors as a precise and complementary cosmological laboratory, alongside their overdense counterparts~\citep{2025arXiv250608932W, 2026arXiv260422105S}.

\section*{Acknowledgments}
During the preparation of this work, KS, EK, WC, and AP were supported in part by grant NSF PHY-2309135 to the Kavli Institute for Theoretical Physics (KITP). This material is based upon work supported in part by the National Science Foundation under CAREER Grant Number 2240032, and the David \& Lucile Packard Foundation grant 2020-71384. WC is supported by the UKRI grant UKRI2424. AP acknowledges support from the European Research Council (ERC) under the European Union’s Horizon programme (COSMOBEST ERC funded project, grant agreement 101078174), and from the French government under the France 2030 investment plan, as part of the Initiative d’Excellence d’Aix- Marseille Université - A*MIDEX AMX-22-CEI-03. We gratefully acknowledge the CHIME/FRB Collaboration for producing the largest FRB catalog published to date that forms the foundation of this work.

\facilities{
CHIME, 
SDSS.
}

\software{
\textsc{Astropy},
\textsc{BaryonForge}, 
\textsc{Cartopy},
\textsc{Pyccl},
\textsc{SciPy},
\textsc{TreeCorr}.
}

\bibliography{manuscript}{}
\bibliographystyle{aasjournal}

\end{document}